\RequirePackage{lineno}
\documentclass[twocolumn,prl,aps,showpacs]{revtex4}

\usepackage{slashed}
\usepackage{graphicx}

\begin{document}

\title{Observation of a Centrality-Dependent Dijet Asymmetry in Lead-Lead Collisions 
at $\sqrt{s_{NN}}=2.76$ TeV with the ATLAS Detector at the LHC}

\author{G. Aad \textit{et al.} (The ATLAS Collaboration)}\thanks{Full author list given at the end of the 
article in Appendix~\ref{app:ATLASColl}.}
\collaboration{The ATLAS Collaboration}

\begin{abstract}
Using the ATLAS detector, observations have been made of a centrality-dependent
dijet asymmetry in the collisions of lead ions at the
Large Hadron Collider.  In a sample of lead-lead events
with a per-nucleon center of mass energy of 2.76 TeV, selected with a 
minimum bias trigger, jets are reconstructed in 
fine-grained, longitudinally-segmented electromagnetic and hadronic calorimeters.
The underlying event is measured
and subtracted event-by-event, giving estimates of jet transverse energy above the ambient
background.  
The transverse energies of dijets in opposite hemispheres is
observed to become systematically more unbalanced with increasing event
centrality leading to a large number of events which contain highly asymmetric
dijets.
This is the first observation of an enhancement of events with such large dijet asymmetries, 
not observed in proton-proton collisions,
which may point to an interpretation in terms of strong jet energy loss in a hot, dense medium.

\end{abstract}
\maketitle


\begin{figure*}[th]
\begin{center}
\includegraphics[width=36pc]{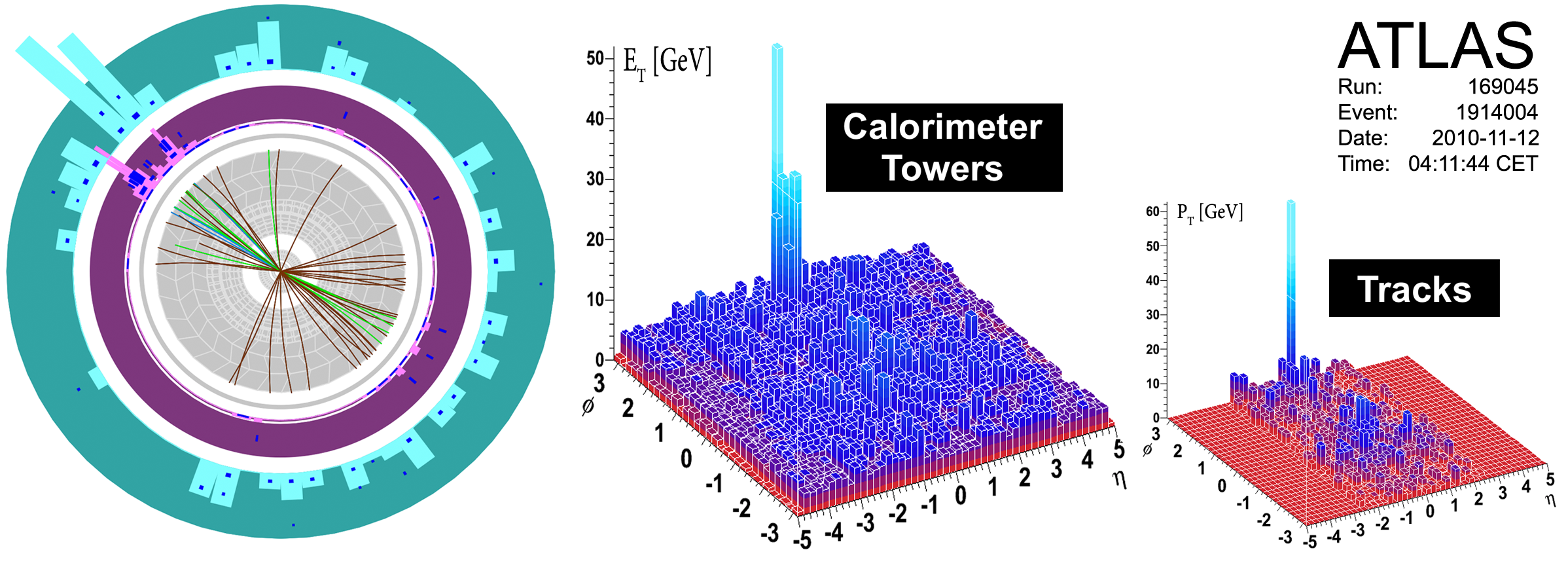}
\caption{\label{event_display}Event display of a highly asymmetric dijet event, 
with one jet with $E_T > 100$~GeV and no evident recoiling jet, and with high energy 
calorimeter cell deposits distributed over a wide azimuthal region.  
By selecting tracks with $p_T>2.6$~GeV and applying cell thresholds in the 
calorimeters ($E_T>700$ MeV in the electromagnetic 
calorimeter, and $E>1$ GeV in the hadronic calorimeter) the recoil can be seen 
dispersed widely over azimuth.}
\end{center}
\end{figure*}

Collisions of heavy ions at ultra-relativistic energies are expected to
produce an evanescent hot, dense state, with temperatures exceeding
two trillion kelvins, in which the relevant degrees of freedom are
not hadrons, but quarks and gluons.
In this medium, high-energy quarks and gluons are expected to transfer energy to the medium
by multiple interactions with the ambient plasma.
There is a rich theoretical literature on in-medium QCD energy loss extending back to Bjorken, who
proposed to look for ``jet quenching'' in proton-proton collisions~\cite{Bjorken:1982tu}.
This work also suggested the observation of highly unbalanced dijets when one jet
is produced at the periphery of the collision.  For comprehensive reviews of recent
theoretical work in this area, 
see Refs.~\cite{Wiedemann:2009sh,Majumder:2010qh}.

Single particle measurements made by RHIC experiments established
that high transverse momentum ($p_T$) hadrons are produced at rates a factor of 
five or more lower than expected
by assuming QCD factorization holds in every binary collision of nucleons in
the oncoming nuclei~\cite{Adcox:2001jp,Adler:2002xw}.  
This observation is characterized by measurements of $R_{AA}$, the ratio of yields in heavy ion collisions
to proton-proton collisions, divided by the number of binary collisions.
Di-hadron measurements also showed a clear absence of back-to-back hadron production
in more central heavy ion collisions~\cite{Adler:2002xw}, 
strongly suggestive of jet suppression.
The limited rapidity coverage of the experiment, and jet energies
comparable to the underlying event energy, prevented a stronger conclusion 
being drawn from these data.

The LHC heavy ion program was foreseen to provide an opportunity to
study jet quenching at much higher jet energies than achieved at RHIC.
This letter provides the first measurements of jet production in
lead-lead collisions at  $\sqrt{s_{NN}}=2.76$ TeV per nucleon-nucleon collision, the
highest center of mass energy ever achieved for nuclear
collisions.  
At this energy, next-to-leading-order QCD 
calculations~\cite{Jager:2004jh} predict abundant rates of jets 
above 100 GeV produced in the pseudorapidity region $|\eta|<4.5$~\cite{refsys}, 
which can be reconstructed by ATLAS.

The data in this paper were obtained by ATLAS during the 2010 lead-lead
run at the LHC
and correspond to an integrated luminosity of approximately 1.7~$\mu b^{-1}$. 

For this study, the focus is on the balance between the
highest transverse energy pair of jets in events where those jets have an
azimuthal angle separation, $\Delta \phi =|\phi_1-\phi_2| > \pi/2$ to reduce
contributions from multi-jet final states.  In this letter, 
jets with $\Delta \phi > \pi/2$ are labeled as being in opposite hemispheres.
The jet energy
imbalance is expressed in terms of the asymmetry $A_J$, 
\begin{equation}
A_J = \frac{E_{T1}-E_{T2}}{E_{T1}+E_{T2}}, \Delta\phi > \frac{\pi}{2}
\end{equation}
where the first jet is required to have a transverse energy $E_{T1} > 100$~GeV,
and the second jet is the highest transverse energy jet in the 
opposite hemisphere with $E_{T2} > 25$~GeV. 
The average contribution of the underlying event energy is subtracted
when deriving the individual jet transverse energies.
The event selection is chosen such that the first jet has high
reconstruction efficiency and the second jet is above the
distribution of background fluctuations and the intrinsic soft jets
associated with the collision. Dijet events are expected to have 
$A_J$ near zero, with deviations expected from 
gluon radiation falling outside the jet cone, as well
as from instrumental effects.  Energy loss in the
medium could lead to much stronger deviations in the reconstructed
energy balance.

The ATLAS detector~\cite{Aad:2008zzm}
is well-suited for measuring jets due to its large acceptance,
highly segmented electromagnetic (EM) and hadronic calorimeters. 
These allow efficient reconstruction of jets over a wide range in the region
$|\eta|<4.5$.  The detector also provides precise charged particle
and muon tracking.  An event display showing the Inner Detector and calorimeter systems
is shown in Fig.~\ref{event_display}.

Liquid argon (LAr) technology providing excellent energy and position
resolution is used in the electromagnetic calorimeter that
covers the pseudorapidity range $|\eta| < 3.2$. The hadronic
calorimetry in the range $|\eta| < 1.7$ is provided by a sampling
calorimeter made of steel and scintillating tiles. In the end-caps
($1.5 < |\eta| < 3.2$), LAr technology is also used for the hadronic
calorimeters, matching the outer $|\eta|$ limits of the
electromagnetic calorimeters. To complete the $\eta$ coverage, the LAr
forward calorimeters provide both electromagnetic and hadronic energy
measurements, extending the coverage up to $|\eta| = 4.9$. 
The calorimeter ($\eta$,$\phi$) granularities are $0.1 \times 0.1$ for the hadronic
calorimeters up to $|\eta| = 2.5$ (except for the third layer of the Tile calorimeter,
which has a segmentation of $0.2 \times 0.1$ up to $|\eta| = 1.7$),
and then $0.2 \times 0.2$ up to $|\eta| = 4.9$.
The EM calorimeters are longitudinally segmented into three
compartments and feature a much finer readout granularity varying
by layer, with cells as small as $0.025 \times 0.025$ extending to
$|\eta| = 2.5$ in the middle layer. 
In the data taking period considered, approximately 187,000 calorimeter cells
(98\% of the total) were usable for event reconstruction.


\begin{figure}[t]
\begin{center}
\includegraphics[width=20pc]{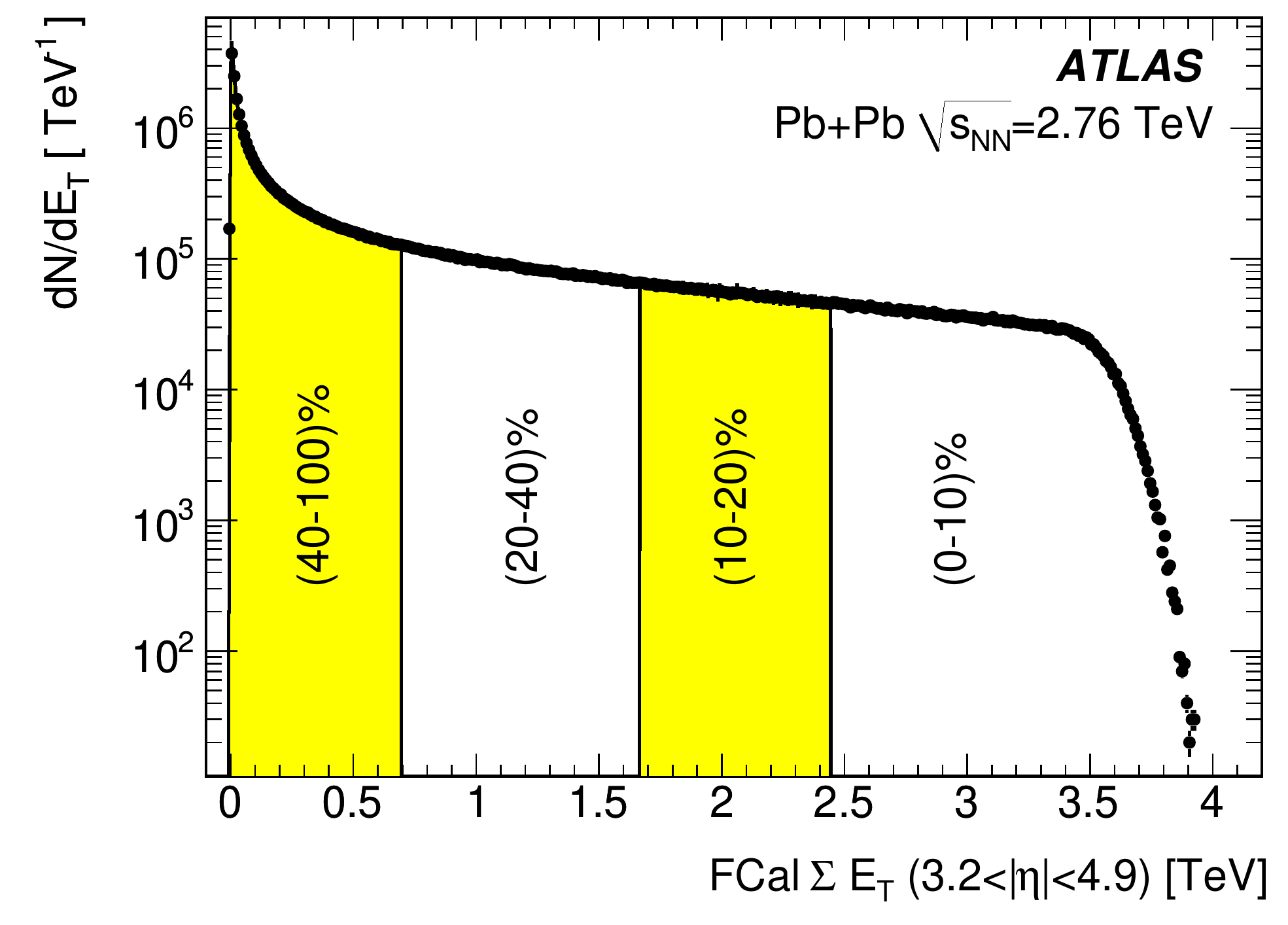}
\includegraphics[width=19.5pc]{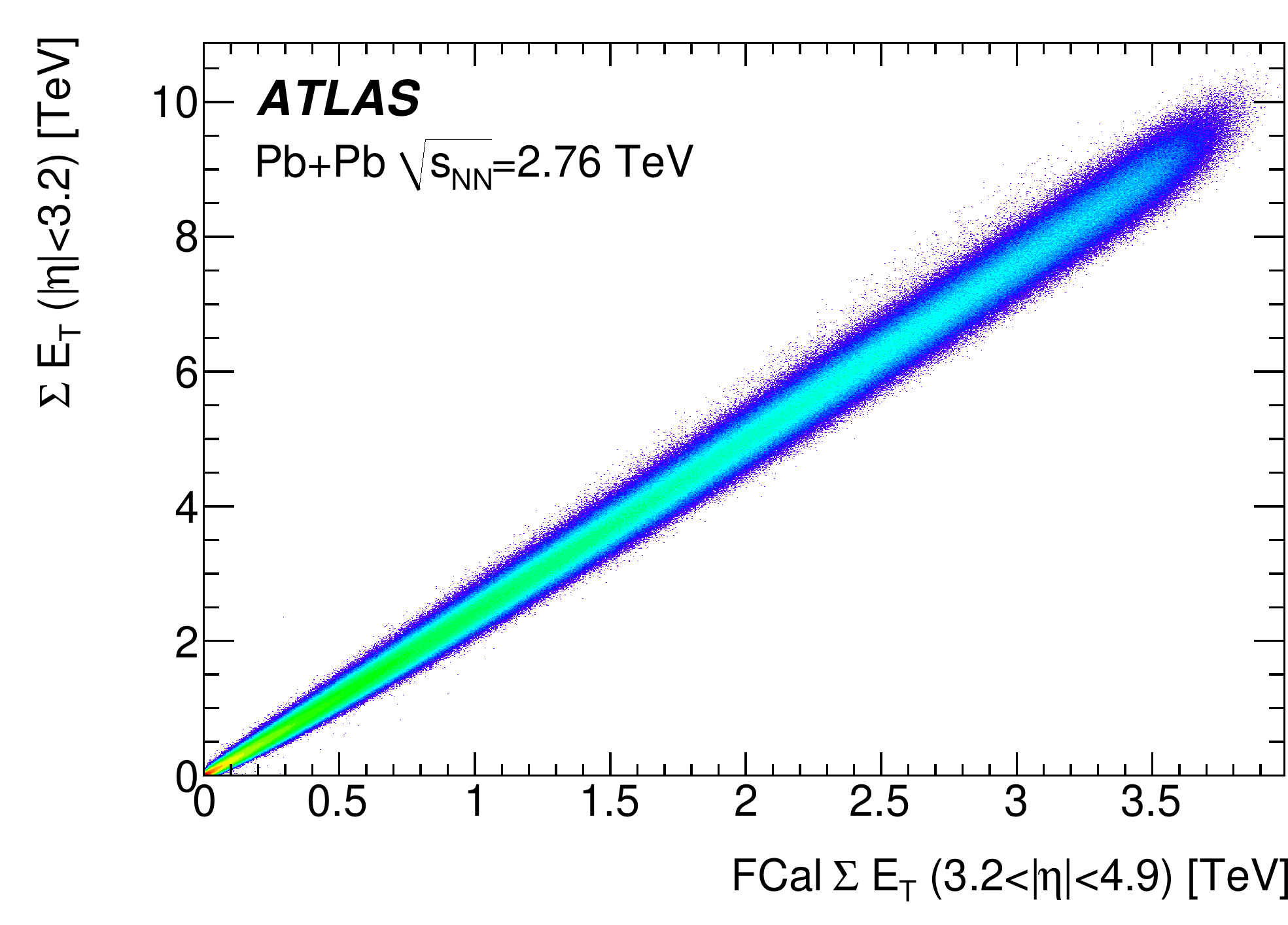}
\caption{\label{EtCentrality}
(top) Distribution of uncorrected $\Sigma E_T$ in the
Forward Calorimeter (FCal). Bins in event activity
or ``centrality'' are indicated by the alternating bands (see text for
details) and labeled according to increasing fraction of lead-lead total
cross section starting from the largest measured $\Sigma E_T$.
(bottom)
Correlation of uncorrected $\Sigma E_T$ in $|\eta|<3.2$ with that
measured in the FCal ($3.2<|\eta|<4.9$).}
\end{center}
\end{figure}
The bulk of the data reported here were triggered 
using coincidence signals from two
sets of Minimum Bias Trigger Scintillator (MBTS) detectors,
positioned at $z = \pm 3.56$~m, covering the full azimuth between
$2.09<|\eta|<3.84$ and divided into eight $\phi$ sectors and two $\eta$
sectors.  Coincidences in the Zero Degree Calorimeter 
and LUCID luminosity detectors were also used as primary triggers, 
since these detectors were far less susceptible to LHC beam
backgrounds.  These triggers have a large overlap and are close to fully
efficient for the events studied here.

In the offline analysis, events are required to have a time difference
between the two sets of MBTS counters of $\Delta t < 3$ ns and a
reconstructed vertex to efficiently reject beam-halo backgrounds.
The primary vertex is derived from the 
reconstructed tracks in the Inner Detector (ID), which covers
$|\eta|<2.5$ using silicon pixel and strip detectors surrounded by
straw tubes.  These event selection criteria have been estimated to
accept over 98\% of the total lead-lead inelastic cross section. 

The level of event activity or ``centrality'' is characterized
using the total transverse energy ($\Sigma E_T$) deposited in the
Forward Calorimeters (FCal), which cover $3.2<|\eta|<4.9$, shown in 
Fig.~\ref{EtCentrality}. Bins are defined in centrality according to
fractions of the total lead-lead cross section selected by the trigger and
are expressed in terms of percentiles (0-10\%, 10-20\%, 20-40\% and
40-100\%) with 0\% representing the upper end of the $\Sigma E_T$
distribution. Previous heavy ion experiments have shown a clear
correlation of the $\Sigma E_T$ with the geometry of the overlap
region of the colliding nuclei and, correspondingly, the total event
multiplicity. This is verified in the bottom panel of
Fig.~\ref{EtCentrality} which shows a tight correlation between the
energy flow near mid-rapidity and the forward $\Sigma E_T$.  
The forward $\Sigma E_T$ is used for this
analysis to avoid biasing the centrality measurement with jets.

Jets have been reconstructed using the infrared-safe anti-$k_t$ jet clustering
algorithm~\cite{Cacciari:2008gp} with the radius parameter $R = 0.4$. 
The inputs to this
algorithm are ``towers'' of calorimeter cells of size $\Delta\eta\times\Delta\phi = 0.1 \times 0.1$
with the input cells weighted using 
energy-density dependent factors to correct for calorimeter non-compensation 
and other energy losses. Jet four-momenta are constructed by the vectorial addition of
cells, treating each cell as an ($E$, $\vec{p}$) four-vector with zero mass.

\begin{figure*}[t]
\begin{center}
\includegraphics[width=40pc]{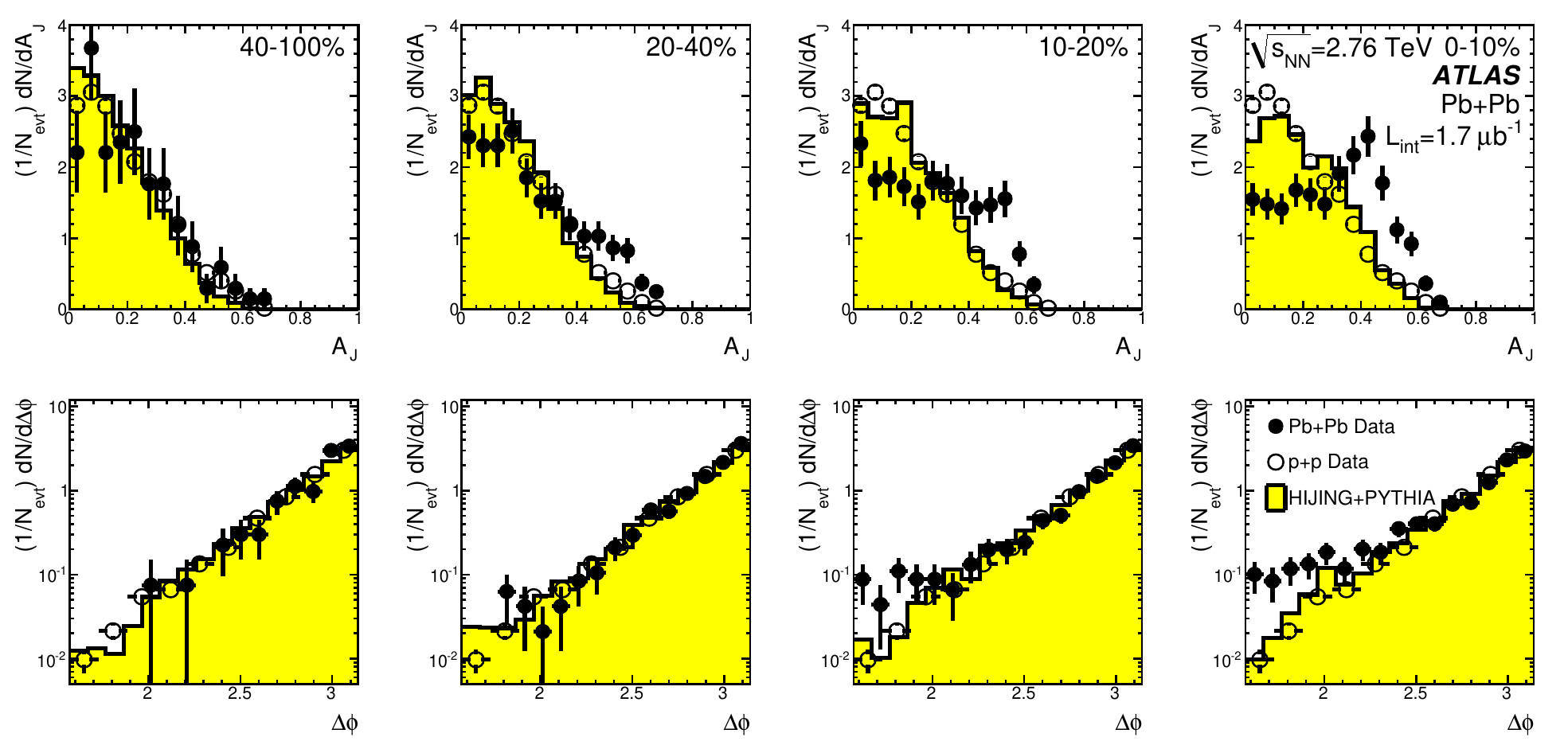}
\caption{\label{final_4x2}
(top) Dijet asymmetry distributions for data (points) and unquenched HIJING with superimposed PYTHIA dijets (solid yellow histograms), 
as a function of collision centrality (left to right from peripheral 
to central events).  Proton-proton data from $\sqrt{s}=7$ TeV, analyzed with the same jet selection, is shown as open circles.
(bottom) Distribution of $\Delta\phi$, the azimuthal angle between the two jets, for data and HIJING+PYTHIA, also as a function of centrality.
}
\end{center}
\end{figure*}

The jets reconstructed using the anti-$k_t$ algorithm contain a mix of 
genuine jets and jet-sized patches of the underlying event.  
For each event, we estimate the average transverse energy density 
in each calorimeter layer in bins of width $\Delta\eta = 0.1$, and 
averaged over azimuth.  In the averaging, we exclude jets with 
$D = E_T(max)/\langle E_T \rangle$,  the ratio of the maximum tower energy over the 
mean tower energy, greater than 5.  The value $D_{cut} = 5$ is chosen 
based upon simulation studies, and the results have been tested 
to be stable against variations in this parameter.  These average 
energies are subtracted layer-by-layer from the cells that make 
up each jet, scaling appropriately for the cell area. The final 
reported four-momentum for each jet is then recalculated from 
the remaining energy in the cells.


The efficiency of the jet reconstruction algorithm, and other event properties, 
have been studied using PYTHIA~\cite{Sjostrand:2006za} events superimposed on 
HIJING events~\cite{Wang:1991hta}. 
There is no parton-level interference between the PYTHIA and HIJING generated events. 
A GEANT4~\cite{Agostinelli:2002hh} simulation models the detector response~\cite{ATLASSim} 
to all the final state particles from the two generated events. 
The HIJING parameters used do not include jet quenching, but variations 
in flow as a function of centrality are added.  It is found that jets with 
$E_T > 100$~GeV are reconstructed   with nearly 100\% efficiency at all centralities.

Simulations have been used to check the overall linearity and resolution of the
reconstruction with respect to the primary jet energy, assuming jet shapes
similar to those found in proton-proton collisions~\cite{Collaboration:2010wv}.  
However, the efficiency, linearity, and resolution for reconstructing jets may be 
poorer if the jets are substantially modified by the medium.  To check the 
sensitivity to such effects, the jet shape,
characterized here as the ratio of the ``core'' energy (integrated
over $\sqrt{\Delta\eta^2+\Delta\phi^2}<0.2$) to the total energy, has been studied.  This ratio
shows only a weak dependence on centrality, providing
evidence that the high-energy jets do look approximately like
jets measured in proton-proton collisions, and that the energy subtraction
procedure does not introduce significant biases.


After event selection, the requirement of a leading jet with $E_T>100$
GeV and $|\eta| < 2.8$  yields a sample of 1693 events. These are called the ``jet
selected events''. 
The lead-lead data are also compared with a sample of 17 $nb^{-1}$ 
of proton-proton collision data~\cite{Collaboration:2010wv}, which yields 6732 events.

A striking feature of this sample is the appearance of events with
only one high $E_T$ jet clearly visible in the calorimeter, and no
high $E_T$ jet opposite to it in azimuth. Such an event is shown in
Fig.~\ref{event_display}.
The calorimeter $E_T$ and charged particle $\Sigma
p_T$ are shown in regions of $\Delta \eta \times \Delta \phi = 0.1\times 0.1$.
Inspection of this event shows a highly asymmetric pair
of jets with the particles recoiling against the leading jet being
widely distributed in azimuth.

To quantify the transverse energy balance between jets in these events, we
calculate the dijet asymmetry, $A_J$, in different
centrality bins between the highest $E_T$ (leading) jet and the
highest $E_T$ jet in the opposite hemisphere (second jet). 
The second jet is required to have $E_T >$ 25 GeV in order to discriminate against background from the 
underlying event.  This excludes around 5\% of the jet selected events in the most
central 40\% of the cross section, and accepts nearly all of the more peripheral events.
 
The dijet asymmetry and $\Delta\phi$ distributions are shown in four centrality
bins in Fig.~\ref{final_4x2}, where they are compared with proton-proton data and with
fully-reconstructed HIJING+PYTHIA simulated events.
The simulated events are intended to illustrate the effect of the
heavy ion background on jet reconstruction, not any underlying physics
process. The dijet asymmetry in peripheral lead-lead events
is similar to that in both proton-proton and simulated events; however, as
the events become more central, the lead-lead data distributions develop different
characteristics, indicating an increased rate of highly asymmetric dijet
events. The asymmetry distribution broadens; the mean shifts to higher values;
the peak at zero asymmetry is no longer visible; and for the most central
events a peak is visible at higher asymmetry values (asymmetries larger than
0.6 can only exist for leading jets substantially above the kinematic
threshold of 100 GeV transverse energy). The $\Delta\phi$ distributions show that
the leading and second jets are primarily back-to-back in all centrality bins;
however, a systematic increase is observed in the rate of second jets at large
angles relative to the recoil direction as the events become more central.

Numerous studies have been performed to verify that the events with large 
asymmetry are not produced by backgrounds or detector effects.
Detector effects primarily include readout errors and local acceptance loss due to 
dead channels and detector cracks. All of the jet events in this sample were checked,
and no events were flagged as problematic.
The analysis was repeated first requiring both jets to be within $|\eta|<1$ and $|\eta|<2$, to see if there
is any effect related to boundaries between the calorimeter sections, and no change to the distribution
was observed. Furthermore, the highly-asymmetric dijets were not found to populate 
any specific region of the calorimeter, indicating that no substantial fraction of 
produced energy was lost in an inefficient or uncovered region.

To investigate the effect of the underlying event, the
jet radius parameter $R$ was varied from 0.4 to 0.2 and 0.6 with the result 
that the large asymmetry was not reduced. 
In fact, the asymmetry increased for the smaller radius, which would not be expected 
if detector effects are dominant. The analysis was independently corroborated by a study 
of ``track jets'', reconstructed with ID tracks of $p_T>4$ GeV using the same jet algorithms.  
The ID has an estimated efficiency for reconstructing charged hadrons above $p_T>1$~GeV of approximately
80\% in the most peripheral events (the same as that found in 7~TeV
proton-proton operation) and 70\% in the most central events, due to the
approximately 10\% occupancy reached in the silicon strips. 
A similar asymmetry effect is also observed with track jets. 
The jet energy scale and underlying event 
subtraction were also validated by correlating calorimeter and track-based jet measurements.

The missing $E_T$ distribution was measured for minimum bias heavy ion events as a function of the
total $E_T$ deposited in the calorimeters up to about $\Sigma E_T = 10$~TeV.  
The resolution as a function of total $E_T$ shows the same behavior as
in proton-proton collisions. 
None of the events in the jet selected sample was found to have an anomalously 
large missing $E_T$.

The events containing high-$p_T$ jets were studied for the presence of high-$p_T$ muons that could carry 
a large fraction of the recoil energy.  Fewer than 2\% of the events have a muon with $p_T > 10$~GeV, potentially
recoiling against the leading jet, so this can not explain the prevalence of highly asymmetric 
dijet topologies in more central events.

None of these investigations indicate that the highly-asymmetric dijet events 
arise from backgrounds or detector-related effects.

In summary, first results are presented on jet reconstruction in lead-lead collisions, with
the ATLAS detector at the LHC. In a sample of events with a reconstructed jet 
with transverse energy of 100 GeV or more, 
an asymmetry is observed between the transverse energies of the leading and
second jets that increases with the centrality of the collisions.
This has a natural interpretation in terms of QCD energy loss, where
the second jet is attenuated, in some cases leading to striking highly-asymmetric 
dijet events.
This observation is the first of an enhancement of such large dijet asymmetries, 
not observed in proton-proton collisions,
which may point to an interpretation in terms of strong jet energy loss in a hot, dense medium.

We wish to thank CERN for the efficient commissioning and operation of the LHC during this 
initial high-energy data-taking period as well as the support staff from our institutions 
without whom ATLAS could not be operated efficiently.We acknowledge the support of ANPCyT, 
Argentina; YerPhI, Armenia; ARC, Australia; BMWF, Austria; ANAS, Azerbaijan; SSTC, Belarus; CNPq and 
FAPESP, Brazil; NSERC, NRC and CFI, Canada; CERN; CONICYT, Chile; CAS, MOST and NSFC, China; COLCIENCIAS, 
Colombia; MEYS (MSMT), MPO and CCRC, Czech Republic; DNRF, DNSRC and Lundbeck Foundation, Denmark; ARTEMIS, 
European Union; IN2P3-CNRS, CEA-DSM/IRFU, France; GNAS, Georgia; BMBF, DFG, HGF, MPG and 
AvH Foundation, Germany; GSRT, Greece; ISF, MINERVA, GIF, DIP and Benoziyo Center, Israel; 
INFN, Italy; MEXT and JSPS, Japan; CNRST, Morocco; FOM and NWO, Netherlands; RCN, Norway;  
MNiSW, Poland; GRICES and FCT, Portugal;  MERYS (MECTS), Romania;  MES of Russia and ROSATOM, 
Russian Federation; JINR; MSTD, Serbia; MSSR, Slovakia; ARRS and MVZT, Slovenia; DST/NRF, 
South Africa; MICINN, Spain; SRC and Wallenberg Foundation, Sweden; SER,  
SNSF and Cantons of Bern and Geneva, Switzerland;  NSC, Taiwan; TAEK, Turkey; STFC, 
the Royal Society and Leverhulme Trust, United Kingdom; DOE and NSF, United States of America. 
The crucial computing support from all WLCG partners is acknowledged gratefully, in 
particular from CERN and the ATLAS Tier-1 facilities at TRIUMF (Canada), 
NDGF (Denmark, Norway, Sweden), CC-IN2P3 (France), KIT/GridKA (Germany), 
INFN-CNAF (Italy), NL-T1 (Netherlands), PIC (Spain), ASGC (Taiwan), 
RAL (UK) and BNL (USA) and in the Tier-2 facilities worldwide.


\begin{thebibliography}{15}

\bibitem{Bjorken:1982tu}
J.~D.~Bjorken,
FERMILAB-PUB-82-059-THY (1982).

\bibitem{Wiedemann:2009sh}
 U.~A.~Wiedemann,
 arXiv:0908.2306 [hep-ph].

\bibitem{Majumder:2010qh}
  A.~Majumder and M.~Van Leeuwen,
  arXiv:1002.2206 [hep-ph].

\bibitem{Adcox:2001jp}
 K.~Adcox et al. [PHENIX Collaboration],
Phys. Rev. Lett. {\bf 88}, 022301 (2002)

\bibitem{Adler:2002xw}
 C.~Adler {\it et al.}  [STAR Collaboration],
 Phys.\ Rev.\ Lett.\  {\bf 89}, 202301 (2002)

\bibitem{Jager:2004jh}
 B.~Jager, M.~Stratmann and W.~Vogelsang,
 Phys.\ Rev.\  D {\bf 70}, 034010 (2004)

\bibitem{refsys}
The ATLAS reference
system is a Cartesian right-handed coordinate system, with the nominal
collision point at the origin. The anticlockwise beam direction defines
the positive z-axis, while the positive x-axis is defined as pointing from
the collision point to the center of the LHC ring and the positive y-axis
points upwards. The azimuthal angle $\phi$ is measured around the beam axis,
and the polar angle $\theta$ is measured with respect to the z-axis.   
Pseudorapidity is defined as $\eta = - \ln(\tan(\theta/2))$. 

\bibitem{Aad:2008zzm}
 The ATLAS Collaboration, 
 JINST {\bf 3}, S08003 (2008).

\bibitem{Cacciari:2008gp}
 M.~Cacciari, G.~P.~Salam and G.~Soyez,
 JHEP {\bf 0804}, 063 (2008)

\bibitem{Sjostrand:2006za}
 T.~Sjostrand, S.~Mrenna and P.~Z.~Skands,
 JHEP {\bf 0605}, 026 (2006)

\bibitem{Wang:1991hta}
X.-N. Wang and M. Gyulassy,
Phys. Rev. {\bf D44}, 3501 (1991).

\bibitem{Agostinelli:2002hh}
 S.~Agostinelli {\it et al.}  [GEANT4 Collaboration],
 Nucl.\ Instrum.\ Meth.\  A {\bf 506}, 250 (2003).
 
\bibitem{ATLASSim}
 The ATLAS Collaboration,
 http://dx.doi.org/10.1140/epjc/s10052-010-1429-9 Eur. Phys. J. C (2010)


\bibitem{Collaboration:2010wv}
  The ATLAS Collaboration, 
  arXiv:1009.5908 [hep-ex], Accepted by EPJC.

\end{thebibliography}

\clearpage
\onecolumngrid
\section{The ATLAS Collaboration}
\label{app:ATLASColl}
\begin{center}
\begin{flushleft}

\bigskip

G.~Aad$^{\rm 48}$,
B.~Abbott$^{\rm 111}$,
J.~Abdallah$^{\rm 11}$,
A.A.~Abdelalim$^{\rm 49}$,
A.~Abdesselam$^{\rm 118}$,
O.~Abdinov$^{\rm 10}$,
B.~Abi$^{\rm 112}$,
M.~Abolins$^{\rm 88}$,
H.~Abramowicz$^{\rm 153}$,
H.~Abreu$^{\rm 115}$
% ...
\bigskip

% ...
$^{10}$ Institute of Physics, Azerbaijan Academy of Sciences, H. Javid Avenue 33, AZ 143 Baku, Azerbaijan\\
$^{11}$ Institut de F\'isica d'Altes Energies, IFAE, Edifici Cn, Universitat Aut\`onoma  de Barcelona,  ES - 08193 Bellaterra (Barcelona), Spain\\
% ...
$^{48}$ Albert-Ludwigs-Universit\"{a}t, Fakult\"{a}t f\"{u}r Mathematik und Physik, Hermann-Herder Str. 3, D - 79104 Freiburg i.Br., Germany\\
$^{49}$ Universit\'e de Gen\`eve, Section de Physique, 24 rue Ernest Ansermet, CH - 1211 Geneve 4, Switzerland\\
% ...
$^{88}$ Michigan State University, Department of Physics and Astronomy, High Energy Physics Group, East Lansing, MI 48824-2320, United States of America\\
% ...
$^{111}$ University of Oklahoma, Homer L. Dodge Department of Physics and Astronomy, 440 West Brooks, Room 100, Norman, OK 73019-0225, United States of America\\
$^{112}$ Oklahoma State University, Department of Physics, 145 Physical Sciences Building, Stillwater, OK 74078-3072, United States of America\\
% ...
$^{115}$ LAL, Univ. Paris-Sud, IN2P3/CNRS, Orsay, France\\
% ...
$^{118}$ Oxford University, Department of Physics, Denys Wilkinson Building, Keble Road, Oxford OX1 3RH, United Kingdom\\
% ...
$^{153}$ Tel Aviv University, Raymond and Beverly Sackler School of Physics and Astronomy, Ramat Aviv, IL - Tel Aviv 69978, Israel\\
% ...
\end{flushleft}


\end{center}

\end{document}